\documentclass[
 reprint,
bibnotes,
 amsmath,amssymb,
 aps,
prb,
]{revtex4-2}

\usepackage{wrapfig}
\usepackage{graphicx}
\usepackage{dcolumn}
\usepackage{bm}
\usepackage{hyperref}
\usepackage[mathlines]{lineno}

\usepackage[utf8]{inputenc}
\usepackage{amsmath,mleftright}
\usepackage{array}
\usepackage{multirow}
\usepackage[margin=1in]{geometry}

\usepackage{color}
\usepackage{colortbl}
\usepackage{physics}

\begin{document}

\definecolor{red}{rgb}{1,0,0}
\definecolor{green}{rgb}{0,1,0}
\definecolor{yellow}{rgb}{1,1,0}
\definecolor{blue}{rgb}{.36,.54,.66}
\definecolor{amethyst}{rgb}{0.6, 0.4, 0.8}
\definecolor{orange}{rgb}{1,.5,0}
\definecolor{fuchsia}{rgb}{.57,.36,.51}
\definecolor{tangerine}{rgb}{1,.6,.4}

\preprint{Crap}

\title{Proposal for methods to measure the octupole susceptibility}

\author{M.E. Sorensen$^{1,2}$ and I.R. Fisher$^{1,3}$}
\affiliation{1. Geballe Laboratory for Advanced Materials, Stanford University, California 94305, USA}
\affiliation{2. Department of Physics, Stanford University, Stanford, CA 94305}
\affiliation{3. Department of Applied Physics, Stanford University, Stanford, CA 94305}

\date{\today}

\begin{abstract}
Direct means of measuring the susceptibility towards an octupole order parameter are proposed via a sixth-rank tensor property. Equivalent derivatives of more conventionally measured tensor properties, including elastic stiffness, magnetic susceptibility, and elastoresistivity, are written in full, as constrained by the symmetry of the experimentally-motivated $O_h$ point group. For simplicity, we consider the specific case of $Pr^{3+}$ ions in a cubic point symmetry with a $\Gamma_3$ crystal field ground state, but the ideas are somewhat general. Experimental feasibility of measuring these various derivatives of tensor quantities is discussed.
\end{abstract}

\maketitle

\section{Introduction}

Quantum states of localized electrons can have a variety of well-defined electromagnetic multipole moments; indeed, within higher angular-momentum states, particularly those that often arise from f-orbitals, higher-order multipolar moments frequently have some non-zero expectation value in the presence of simple magnetic (dipole) order. Of course, multipolar moments are subject to higher-order interactions amongst themselves, and can thus order independently of any dipole moment, but this is uncommon: the dipole typically dominates in energy scale whenever a variety of multipoles are present or allowed. This motivates the use of the associated multipole susceptibility as a powerful tool for analyzing these higher-order multipoles, as the strength and character of specific multipolar interactions can be probed without requiring a simple ordered state of such a multipole. For higher-rank multipoles, however, it is a non-trivial task to relate the multipole susceptibility to physically measurable quantities.

It is well established that a ($q=0$) magnetic susceptibility may be measured via an applied uniform magnetic field. Specifically, a magnetic field couples bilinearly to the magnetization (magnetic dipole moments per unit volume), and hence is an appropriate conjugate field. Similarly, antisymmetric strain couples bilinearly to electric quadrupoles, providing access to the quadrupole strain susceptibility [\onlinecite{Elliott}]. Here, we focus on the magnetic octupole, the next in the multipole series \footnote{The electric dipole, magnetic quadrupole, and electric octupole break inversion symmetry, and are thus less common: Crystalline Electric Field ground states that preserve inversion do not include these multipoles as degrees of freedom, leaving them often (but not always) high-energy excited states in 4f systems, rather than potentially ordered ground states}.

Magnetic octupole order has been proposed for many f-orbital systems, but is often hard to verify or probe directly[\onlinecite{Generic1,Generic2,Generic3}]. Given the time-reversal symmetry breaking inherent in a magnetic moment (of any rank), a bilinear coupling of an octupole to strain, like that of the electric quadrupole, is not possible; similarly, a magnetic field will couple bilinearly only to the magnetic dipole moments, with symmetry forbidding a bilinear octupole coupling. As shown in the work of A.S. Patri et. al. [\onlinecite{YongBaek}], however, the combination of the two provides a conjugate field which, by symmetry, can couple directly to the octupole, allowing one to define and measure a susceptibility. This susceptibility most naturally manifests itself in a 6th-rank tensor, in contrast to the 2nd-rank (magnetic susceptibility) tensor for dipoles and the various 4th-rank tensor components representative of quadrupole susceptibilities; the octupole susceptibility can thus be measured independently of the behavior of the lower-order multipoles in the system, at least under certain restrictions.

Measurement of an octupole susceptibility is then possible whenever the specifics of the system render it finite, but possibly quite difficult if lower-order multipoles are present. In particular, lower-order terms in a material's tensor properties invoking the strain or magnetic field individually could potentially drown out any higher-order effects associated with the octupolar degrees of freedom. Thus, the use of Neumann’s principle \footnote{Neumann's principle essentially states that a crystal's physical properties must be invariant under the symmetry operations of the crystal} to significantly constrain the symmetry-allowed tensor terms is motivated, as terms can potentially be identified which have fewer lower-order components or other possible experimental impediments.

Furthermore, we choose to restrict our focus to intermetallic compounds with $Pr^{3+}$ ions in a cubic point symmetry, and for which the crystalline electric field ground state is a $\Gamma_3$ doublet. The specific symmetries of this system forbid all magnetic dipoles and three of the five electric quadrupoles, allowing the octupole's conjugate field to be applied without inducing any lower-order multipoles. This system has been theoretically shown to allow and potentially favor an octupole order parameter [\onlinecite{YongBaek}],  and has been experimentally shown to order in a manner suggestive of an octupole order parameter [~\onlinecite{YongBaek},~\onlinecite{Nakatsuji}], making a measurement of the octupole susceptibility almost certainly feasible.

Thus, herein we propose and elucidate upon the measurement of various tensor components to identify the associated susceptibility of a given octupole, separating it from the susceptibilities of other multipoles and probing interaction strengths of the octupole directly.

\section{Background}

\subsection{Introduction to the $O_h$ Point Group}

While any cubic point group can give rise to a $\Gamma_3$ doublet ground state, the most prominent experimentally realized case for an octupole order parameter has an $O_h$ point group [\onlinecite{Nakatsuji},\onlinecite{Crystal}]. The $O_h$ point group, being the most highly symmetric cubic point group, contains 48 symmetry elements, many of which are redundant in constraining the various tensor properties (see Appendix A for full list of symmetries via a character table). A convenient, less redundant basis to work in is then shown in Table \ref{tab:table1}, where $\sigma_i$ represents a mirror plane defined by the $i$ axis, $C_{xi}$ represents an x-fold rotation about the $i$ axis, and $F_{ij}$ represents some generic 2nd-rank material property tensor.

\begin{table}
\caption{\label{tab:table1}Effects of various mirror planes ($\sigma$) and rotations ($C$) contained in $O_h$ on a generic 2nd-rank tensor}
\begin{tabular}{ |c|c|c|c| } 
\hline
Symmetry & Effect(s) & Implied Equality \\
\hline
$\sigma_{i}$ & $i\rightarrow -i$ & $F_{ij} = -F_{ij}(=0)$ \\ 
$\sigma_{i=\pm j}$ & $i\rightarrow \mp j\rightarrow i$ & $F_{ij} = F_{ji}$ \\
$C_{4k}$ & $i\rightarrow j\rightarrow -i$ & $F_{ij} = -F_{ji}$ \\
$C_{3(111)}$ & $i\rightarrow j\rightarrow k\rightarrow i$ & $F_{ij} = F_{jk}$ \\
\hline
\end{tabular}
\end{table}

These symmetry elements of the point group place constraints on tensor properties of the material via Neumann's principle: the tensor properties must be invariant under the symmetry operations of the point group. In the absence of perturbative fields, these are calculated trivially by applying the symmetries to a given tensor element and observing how they affect the various indices; for example, under $C_{3(111)}$ rotation, $x\rightarrow y$, so 
\begin{eqnarray}
F_{xx} = C_{3(111)} F_{xx} = F_{yy}\nonumber\\
F_{xx}=F_{yy}
\end{eqnarray}
Additional examples are shown in Table \ref{tab:table1}. The presence of additional perturbative fields, such as magnetic field in the elastic tensors or elastic strains in the magnetic susceptibility, breaks the symmetries of the material and allows otherwise forbidden terms. This can be accounted for by simply incorporating the symmetry transformations of the perturbative fields [\onlinecite{Max}].The symmetries of the strain tensor and magnetic field are then relevant to all other tensors, and worth some brief discussion. The strain tensor is defined in a manifestly symmetric manner,
\begin{eqnarray}
\epsilon_{ij}\equiv\frac{\partial\mu_i}{\partial x_j}+\frac{\partial\mu_j}{\partial x_i}
\end{eqnarray}
where $\mu_i$ represents the displacement of an atom along the $i$ axis from the unstrained position $x_i$. The inherent symmetry of the strain tensor then requires $\epsilon_{ij} = \epsilon_{ji}$, but otherwise elements of the strain tensor will transform similarly to any other tensor, via applications of the symmetry operations to their indices \footnote{Herein, all proposed measurements treat strain as an extrinsic property, controllable via application of external force; as such, it is not constrained by Neumann's principle}. Magnetic field, on the other hand, is a pseudovector, invariant under inversion; thus, it transforms as expected under the various rotations, but under mirror planes, which can be considered as a combined rotation and inversion, it effectively experiences only the rotation. Hence, $\sigma_x (H_x, H_y, H_z)$ yields $(H_x, -H_y, -H_z)$, for example, in contrast to an arbitrary normal vector $\sigma_x (A_x, A_y, A_z) = (-A_x, A_y, A_z)$. One can see the effect of these external fields with a brief example: without magnetic field, for instance, one sees

\begin{eqnarray}
F_{xy} = C_{4z} F_{xy} = - F_{yx} \nonumber
\\
F_{xy} = -F_{yx} \nonumber
\\
\sigma_{x=y} F_{xy} = \sigma_{x=y}(-F_{yx}) = -F_{xy} \nonumber
\\
F_{xy} = -F_{xy} = 0 
\end{eqnarray}
i.e. $\chi_{xy}$ (and, by similar symmetries, all $\chi_{ij}$ terms for $i\neq j$) is constrained to be 0.
However, introducing a magnetic-field dependency to these terms yields
\begin{eqnarray}
C_{4z}F_{xy}(H_z) = - F_{yx}(H_z)\nonumber
\\
\sigma_{x=y}(-F_{yx}(H_z)) = -F_{xy}(-H_z) \nonumber
\\
\sigma_{x=y}C_{4z}F_{xy}(H_z) = F_{xy}(H_z) = -F_{xy}(-H_z)
\end{eqnarray}

Thus, $F_{xy}$ is no longer constrained to be zero, but merely constrained to be odd in $H_z$, the external field that breaks the symmetry ($\sigma_{x=y}$) that constrained it to be zero. Terms constrained to be equal in the absence of perturbative fields can have dependencies in fields with slight variation in sign and ordering, but will maintain identical sets of coefficients; for example, while $F_{xy} = F_{yz}$ without field, $F_{yz}(H_x)$ need not be identical $F_{xy}(H_x)$, but must instead be identical to $F_{xy}(H_z)$ (via $C_{3(111)}$), leaving the two terms with identical, if differently ordered, sets of coefficients. Similarly, using the above example, $F_{xy}(H_z) = -F_{yx}(H_z)$, implying $F_{xy}$ and $F_{yx}$ will have the same linear $H_z$ coefficients, but with opposite sign. These symmetry principles will be used in Section III to determine allowed terms in several higher rank tensors. Complete descriptions of how these symmetries apply to the various tensors examined in the text can be found in Appendix B.

\subsection{The $\Gamma_3$ Doublet}

While strong spin-orbit coupling among local 4f electrons often makes $J$ a good quantum number, the crystalline electric field (CEF) splitting in 4f materials can substantially reduce the number of available states within a given $J$ multiplet, at least in a low-temperature regime. One of these CEF eigenstates, the $\Gamma_3$ doublet, is generally present in cubic systems, but is rarely the ground state, meaning it cannot often be experimentally isolated. However, calculations have shown that in the special case $J=4$, associated with the $Pr^{3+}$ ion (with $4f^2$ orbital) \footnote{Using the Russell-Saunders coupling scheme, $U^{4+}$ ($5f^2)$ can also manifest a J=4 state. However, the extended nature of $5f$ orbitals often smears out the CEF eigenstates. Furthermore, in some cases j-j coupling is more appropriate. Hence, $Pr^{3+}$ is the clearest manifestation of a J=4 state}, the doublet is a potential ground state [\onlinecite{Lea}].

The $Pr^{3+}$ ions in the most prominent octupole case exist on a diamond lattice [\onlinecite{Crystal}], so the symmetry of the CEF eigenstates is determined by the $T_d$ point group, as this is the local symmetry an individual ion experiences. The $\Gamma_3$ doublet, with basis states (in $J=4$), is then given by,

\begin{eqnarray}
\Gamma_3^{(1)} = \frac{1}{2}(\sqrt{\frac{7}{6}}\ket{4}-\sqrt{\frac{5}{3}}\ket{0}+\sqrt{\frac{7}{6}}\ket{-4})\nonumber
\\
\Gamma_3^{(2)} = \frac{1}{\sqrt(2)}(\ket{2}+\ket{-2})
\end{eqnarray}

As a two-state space, this can be treated as a pseudo-spin [\onlinecite{YongBaek}], and analogously three operators can potentially split the doublet and create a finite order parameter. Group theory decomposition of the doublet in $T_d$ suggests the symmetry of the allowed operators:
\begin{eqnarray}
\Gamma_3 \otimes \Gamma_3 = \Gamma_3 \oplus \Gamma_2 \oplus \Gamma_1
\end{eqnarray}
Thus, of the three operators that would act as Pauli matrices in this pseudo-half-spin two-state space, two have the symmetry of $\Gamma_3$ ($E$) and one has the $\Gamma_2$ ($A_2$) symmetry. One might thus expect one of these operators to break time-reversal symmetry analogously to the Pauli $S_y$ matrix, and indeed the lowest-order multipole of $\Gamma_2$ symmetry is then time-reversal odd. Thus, from the angular-momentum operators $J_x,J_y,J_z$ and their various products (the Stevens operators), the allowed order parameters are represented by two time-reversal-even quadrupole operators of $\Gamma_3$ symmetry
\begin{eqnarray}
O_2^2 = \frac{\sqrt{3}}{2}(J_x^2-J_y^2) \nonumber\\
O_2^0 = \frac{1}{2}(2J_z^2-J_x^2-J_y^2)
\end{eqnarray}
and one time-reversal-odd octupole operator of $A_2$ symmetry
\begin{eqnarray}
\tau_{xyz} = \frac{\sqrt{15}}{6}\overline{J_xJ_yJ_z}
\end{eqnarray}
where $\overline{J_xJ_yJ_z}$ denotes all permutations of the indices $x,y,z$, i.e. a six-term object.
In typical pseudo-spin fashion, one can note the eigenstates of the three operators in the aforementioned basis: $\Gamma_3^{(1)}$ and $\Gamma_3^{(2)}$ for $O^0_2$, $\Gamma_3^{(1)} \pm \Gamma_3^{(2)}$ for $O^2_2$, and $\Gamma_3^{(1)} \pm i \Gamma_3^{(2)}$ for $\tau_{xyz}$.
It can then be noted that none of these three operators represent and/or commute with a magnetic dipole operator. Indeed, in $T_d$ and other cubic point groups, magnetic dipoles belong to a triply-degenerate $\Gamma_4$ ($T_1$) irreducible representation, an object that, as seen in the group theory decomposition, one cannot construct from the two $\Gamma_3$ ($E$) basis states. More intuitively, this can be explained by the $\Gamma_3$ doublet basis states $\Gamma_3^{(1)}$ and $\Gamma_3^{(2)}$ both having three (primary-axis) $C_2$ rotational symmetries, which are universally broken by a dipole order parameter. Thus, cubic praseodymium compounds are of particular interest in the study of higher-order multipoles, as they provide the opportunity to directly probe time-reversal-odd octupolar signatures without (magnetic) dipole signatures; dipole moments are forbidden, to the extent that the energy separation between the $\Gamma_3$ CEF ground state and any triplet excited states is large relative to the temperature and/or magnetic field.

\subsection{Defining an Octupole Susceptibility}
Given the presence of a potential octupolar moment, the natural question is how best to access it experimentally. As was noted by A.S. Patri et. al. [\onlinecite{YongBaek}], an octupolar susceptibility can easily be defined for a variety of potential order parameters. Here we choose to focus on a $q=0$ order parameter, as this presents the most experimentally accessible possibility. It is also, however, of interest for a broader set of potential order parameters; analogously to the magnetic case, finite-q octupole order parameters would likely appear via a sharp feature of some kind in the $q=0$ octupole susceptibility at or near the relevant ordering temperature.

Based on the symmetry properties of the $\tau_{xyz}$ octupole, one can quickly note that a time-reversal-odd conjugate field would be necessary to couple to it. Utilizing two experimentally-common external fields, strain and magnetic field, it can couple bilinearly to two objects, $H_x\epsilon_{yz} + H_y\epsilon_{zx} + H_z\epsilon_{xy}$ and $H_xH_yH_z$\footnote{Looking to a character table quickly shows these two objects to be of $A_2$ symmetry in $T_d$, when $H$ is properly acknowledged as a pseudovector} (here these are considered uniform, but a finite-q order parameter could be coupled to via similar but staggered fields). Choosing to focus on the former for the moment, one might then expect application of this field
\begin{eqnarray}
H\epsilon \equiv H_x\epsilon_{yz} + H_y\epsilon_{zx} + H_z\epsilon_{xy}
\end{eqnarray}
could induce a finite octupolar moment in an originally unordered state,
\begin{eqnarray}
O \equiv <\tau_{xyz}>
\end{eqnarray}
and one could thus define a susceptibility
\begin{eqnarray}
\chi_O\equiv\frac{\partial O}{\partial(H\epsilon)}
\end{eqnarray}

Here it is worth noting that the octupole has thus far been examined in the $T_d$ point group corresponding to the local symmetry of the $4f$ ion, while discussion on material properties has centered on the $O_h$ point group of the specific material (wherein the $Pr$ sites sit on a diamond lattice [\onlinecite{Crystal}]), which defines the symmetries of the material's tensor properties. Given the chosen coupling field will only induce a ferro-octupolar order parameter, one can note that, while an individual octupole has $\Gamma_2$ symmetry in $T_d$, a pair of aligned octupoles on the two independent ion sites in the greater $O_h$ unit cell correspond to a $\Gamma_2^+$ symmetry\footnote{The magnetic octupole is itself invariant under inversion symmetry, while inversion swaps the two independent ion sites in the broader $O_h$ unit cell (equivalent to the two diamond sublattices); thus, if the local octupoles are aligned identically on the two sites, i.e. a ferro-octupolar configuration, the system is invariant under inversion symmetry ($\Gamma_2^+$ in $O_h$ has the same symmetries as $\Gamma_2$ in $T_d$, plus inversion, owing to $T_d$ being a subgroup of $O_h$).}. Thus, for $O_h$ and for a susceptibility as has been described, the order parameter and conjugate field can be more specifically defined as being $\Gamma_2^+$. More broadly, it can be seen that, given the basis states are invariant under inversion (to within an overall phase), and all three operators are similarly invariant, any ferro-aligned $\Gamma_3$ order parameter in $T_d$ will couple as $\Gamma_3^+$ in the larger $O_h$ unit cell \footnote{analogously to the $\Gamma_2$ case, $\Gamma_3^+$ in $O_h$ is equivalent to $\Gamma_3$ in $T_d$ with added inversion symmetry} ($\Gamma_3^-$ objects can couple bilinearly only to non-ferro-aligned $\Gamma_3$ order parameters, which break the inversion symmetry of the larger $O_h$ cell).

\subsection{Basic Landau Theory}

With this $H\epsilon$-type conjugate field, a motivational, simplified model can be established by looking purely at a potential octupolar order parameter. This choice of longitudinal field does leave the aforementioned issues: strain, a second rank tensor, can couple to a quadrupole moment, while magnetic field can couple to a magnetic dipole, leaving any octupole interactions potentially masked. Here we again take advantage of the $\Gamma_3$ doublet: the two $\Gamma_3$ quadrupole moments couple bilinearly only to the two $\Gamma_3$ strains, $\epsilon_{xx}-\epsilon_{yy}$ and $2\epsilon_{zz}-\epsilon_{xx}-\epsilon_{yy}$, while the $\epsilon_{ij}$ strains present in the octupole conjugate field are of $\Gamma_5^+$ type; they can couple to quadrupoles, but only to the $\Gamma_5^+$-type quadrupoles, which are, like the aforementioned magnetic dipoles, forbidden to the extent that the CEF gap is large relative to temperature and strain. Thus, no CEF-allowed multipoles will couple with any of the objects within the octupole conjugate field, allowing one to safely write a lowest-order free energy for just the octupole moment without ignoring any cross-coupling terms not already 'forbidden' by the CEF splitting:

\begin{eqnarray}
F=\frac{a}{2}O^2 - \lambda(H\epsilon)O+ \frac{C_{44}^0}{2}(\epsilon_{xy}^2+\epsilon_{yz}^2+\epsilon_{zx}^2)
\end{eqnarray}

where $a$ is then assumed to be of the standard form $a_0(T-\theta)$, so as to allow for a continuous octupole phase transition, and $C_{44}^0$ is the un-renormalized elastic stiffness.

Assuming the case of a controlled conjugate field \footnote{Strain is a thermodynamic quantity and the material will adapt a value that minimizes the free energy subject to a given set of stresses. However, experimental configurations can be established in which stresses are applied such that given (measured) strains are established. From a thermodynamic perspective, this is equivalent to a Legendre transformation in which strain now becomes a forced (controllable) parameter.}, one can then note that minimizing free energy requires a finite order parameter,
\begin{eqnarray}
O=\frac{\lambda H\epsilon}{a}
\end{eqnarray}

thus allowing one to solve for the octupole susceptibility

\begin{eqnarray}
\chi_O\equiv\left.\frac{dO}{d(H\epsilon)}\right|_{H\epsilon = 0
}=\frac{\lambda}{a_0(T-\theta)}
\end{eqnarray}

Presuming a temperature-independent coupling of the moment and the field $\lambda$, the octupole susceptibility may then follow a simple Curie-Weiss functional form, particularly in systems with a tendency toward an explicit octupole ordering. More generally (i.e. beyond just $\Gamma_3$ doublet cases), this can be taken as the primary proof-of-existence of a measurable octupole susceptibility:  more complicated temperature dependencies will naturally arise from higher-order terms, but can do so both in systems with and without an independent octupole, given other allowed terms invoking other (biquadratic) multipole couplings. Any free-energy term of the form $H^2\epsilon^2$ (after minimization with respect to the various order parameters) must either invoke the octupole or a product of order parameters (a biquadratic dipole-quadrupole coupling, for instance), and thus will have a more complicated lowest-order temperature dependence, excepting coincidental cancellations. The $\Gamma_3$ case is, of course, already simplified by the necessary components of such a composite term, the three magnetic dipoles and the $xy$/$yz$/$zx$ quadrupoles, requiring excited CEF states. A simple $1/(T-\theta)$ dependence in the relevant free energy term is then a reliable indicator of an independent octupolar order parameter, or one with a tendency to order in the absence of competing phases.

One can then note that the octupole susceptibility, to within some proportionality constant, can be extracted by taking appropriate derivatives of the free energy:

\begin{eqnarray}
\left.\frac{\partial^2F}{\partial(H\epsilon)^2}\right|_{H\epsilon \rightarrow 0} = - \frac{\lambda^2}{a}
\end{eqnarray}

This presents the octupolar susceptibility as being proportional to a term in a sixth-rank magneto-elastic tensor. Of course, simpler and similar quantities also present themselves; one can quickly note that a similar quantity (off by a factor of 2) could be found by taking the derivatives separately, and that $\frac{\partial^2F}{\partial\epsilon^2}$ corresponds to the elastic stiffness tensor, while $\frac{\partial^2F}{\partial H^2}$ corresponds to a magnetic susceptibility. We therefore propose measuring the relevant sixth-rank tensor term, and thus the ($q=0$) octupole susceptibility, via field and/or strain derivatives of more commonly measured tensor quantities; in doing so, the more complicated sixth-rank tensor term can be accessed by well-established and understood experimental methods designed for various second- and fourth-rank tensor quantities.

\section{Thermodynamic Tensors}

Thermodynamic quantities, quantities explicitly representative of derivatives of the free energy, are the most direct potential measurements to capture the octupolar susceptibility. Thus, the most obvious tensor quantities involving strain and magnetic field, elastic stiffness and magnetic susceptibility, are herein enumerated.

It should be noted that all tensors herein are general for the $O_h$ point group; while a given term within a tensor may be of specific interest for the octupole here, the allowed and disallowed terms, and their equalities, are a function solely of the point group (and the definitions of the tensors), and not the details of any given system. The symmetric constraints which allow and/or disallow various terms are detailed in Appendix B. Additionally, it should be noted that none of the coefficients are implied to be equal across tensors, with the exception of a handful of identically-labelled coefficients between the elastic stiffness and (strain-dependent) magnetic susceptibility tensors.

\subsection{Elastic Stiffness Tensor}

The elastic stiffness tensor, defined by $C_{ij,kl}\equiv\frac{\partial^2F}{\partial\epsilon_{ij}\partial\epsilon_{kl}}$, represents the stress (i.e. force) necessary to produce a given set of strains in a material. It inherits several symmetries from its definition and that of the strain tensor, $\epsilon_{ij}\equiv\frac{\partial\mu_i}{\partial x_j}+\frac{\partial\mu_j}{\partial x_i}$. Namely, the definition of the strain tensor requires $\epsilon_{ij}$, and thus $C_{ij,kl}$, is invariant under exchange of $i$ and $j$ (or $k$ and $l$), while the definition of $C_{ij,kl}$ requires it be invariant under exchange of $ij$ and $kl$. These taken together motivate the use of compactified Voigt notation rather than a full 9x9 matrix, as many terms are exactly identical to their neighbors in such a full construct (e.g. $C_{xy,xy} = C_{xy,yx} = C_{yx,xy} = C_{yx,yx}$). 

Taking two field derivatives then reconstructs the desired $\chi_O\equiv\frac{\partial O}{\partial(H\epsilon)}\propto\frac{\partial^2 F}{\partial(H\epsilon)^2}$, and thus the field-dependence of the tensor is the primary point of interest. The aforementioned inherent symmetries combined with those of the point group leave 3 independent non-zero terms in the absence of magnetic field, with arbitrary magnetic fields breaking the point-group symmetries and allowing 10 additional independent coefficients (to second order in field), as can be seen in Table \ref{tab:table2}. The $A_5$ (yellow, diagonal boxes) and $D_3$ (blue, off-diagonal boxes) coefficients would then represent the desired direct probe of octupolar susceptibility:
\begin{eqnarray}
A_5 = \frac{\partial^2 F}{\partial^2(H_i\epsilon_{jk})} \propto \chi_O
\\
D_3 = \frac{\partial^2 F}{\partial(H_i\epsilon_{jk})\partial(H_j\epsilon_{ki})} \propto \chi_O
\end{eqnarray}

\begin{table*}
\caption{\label{tab:table2}The full elastic stiffness tensor in $O_h$, in compactified Voigt notation and to second order in magnetic field, color-coded to indicate which terms are identical.}
\begin{tabular}{ |c|c|c|c|c|c|c| } 
\hline
& $xx$ & $yy$ &$zz$&$yz$&$zx$&$xy$\\
\hline
$xx$ & \cellcolor{red}$C_{11}^{(0)} + A_1H_x^2+$ & \cellcolor{green}$C_{12}^{(0)} + A_4H_z^2+$ & \cellcolor{green}$C_{12}^{(0)} + A_4H_y^2+$&\cellcolor{amethyst}&\cellcolor{orange}$B_1H_y$&\cellcolor{orange}$-B_1H_z$ \\ 
& \cellcolor{red}$A_2(H_y^2+H_z^2)$ & \cellcolor{green}$A_3(H_x^2+H_y^2)$ &\cellcolor{green} $ A_3(H_x^2+H_z^2)$&\cellcolor{amethyst}$+D_1H_yH_z$&\cellcolor{orange}$+D_2H_zH_x$&\cellcolor{orange}$+D_2H_xH_y$ \\ 
\hline
$yy$ &\cellcolor{green}$C_{12}^{(0)}  + A_4H_z^2+$ & \cellcolor{red}$C_{11}^{(0)} + A_1H_y^2+$&\cellcolor{green} $C_{12}^{(0)} + A_4H_x^2+$&\cellcolor{orange}$-B_1H_x$&\cellcolor{amethyst}&\cellcolor{orange}$B_1H_z$ \\
&\cellcolor{green}$ A_3(H_y^2+H_x^2)$ & \cellcolor{red}$ A_2(H_z^2+H_x^2)$& \cellcolor{green}$A_3(H_y^2+H_z^2)$&\cellcolor{orange}$+D_2H_yH_z$&\cellcolor{amethyst}$+D_1H_zH_x$&\cellcolor{orange}$+D_2H_xH_y$ \\
\hline
$zz$ &\cellcolor{green}$C_{12}^{(0)}  + A_4H_y^2+$ &\cellcolor{green} $C_{12}^{(0)} + A_4H_x^2+$&\cellcolor{red}$C_{11}^{(0)} + A_1H_z^2+$&\cellcolor{orange}$B_1H_x$&\cellcolor{orange}$-B_1H_y$&\cellcolor{amethyst} \\
&\cellcolor{green} $A_3(H_z^2+H_x^2)$ & \cellcolor{green}$A_3(H_z^2+H_y^2)$&\cellcolor{red}$A_2(H_x^2+H_y^2)$&\cellcolor{orange}$+D_2H_yH_z$&\cellcolor{orange}$+D_2H_zH_x$&\cellcolor{amethyst}$+D_1H_xH_y$ \\
\hline
$yz$ & \cellcolor{amethyst}&\cellcolor{orange}$-B_1H_x$&\cellcolor{orange}$B_1H_x$& \cellcolor{yellow}$C_{44}^{(0)}+A_5H_x^2+$ & \cellcolor{blue} & \cellcolor{blue} \\
& \cellcolor{amethyst}$+D_1H_yH_z$&\cellcolor{orange}$+D_2H_yH_z$&\cellcolor{orange}$+D_2H_yH_z$ & \cellcolor{yellow}$A_6(H_y^2+H_z^2)$ & \cellcolor{blue}$+D_3H_xH_y$ & \cellcolor{blue}$+D_3H_zH_x$\\
\hline
$zx$ &\cellcolor{orange}$B_1H_y$&\cellcolor{amethyst}&\cellcolor{orange}$-B_1H_y$& \cellcolor{blue} &\cellcolor{yellow} $C_{44}^{(0)}+A_5H_y^2+$&\cellcolor{blue} \\ 
& \cellcolor{orange}$+D_2H_zH_x$&\cellcolor{amethyst}$+D_1H_zH_x$&\cellcolor{orange}$+D_2H_zH_x$ & \cellcolor{blue}$+D_3H_xH_y$ &\cellcolor{yellow} $A_6(H_z^2+H_x^2)$ & \cellcolor{blue}$+D_3H_yH_z$\\
\hline
$xy$ & \cellcolor{orange}$-B_1H_z$&\cellcolor{orange}$B_1H_z$&\cellcolor{amethyst} & \cellcolor{blue} & \cellcolor{blue} & \cellcolor{yellow}$C_{44}^{(0)}+A_5H_z^2+$  \\ 
&\cellcolor{orange}$+D_2H_xH_y$&\cellcolor{orange}$+D_2H_xH_y$&\cellcolor{amethyst}$+D_1H_xH_y$ & \cellcolor{blue}$+D_3H_zH_x$  & \cellcolor{blue}$+D_3H_yH_z$ &\cellcolor{yellow} $A_6(H_x^2+H_y^2)$\\
\hline
\end{tabular}
\end{table*}
\subsubsection{Practical Considerations}
A number of considerations present themselves in potential measurements of the relevant coefficients. First, it should be noted that while $A_5$ is unconstrained in its sign by symmetry, the octupole contribution to $A_5$ would necessarily be negative, or correspond to a softening of the lattice:
\begin{eqnarray}
F = -\frac{\lambda^2(H\epsilon)^2}{2a} + \frac{C_{44}^0}{2}(\epsilon_{xy}^2+\epsilon_{yz}^2+\epsilon_{zx}^2) \nonumber\\
C_{44} = \frac{\partial^2 F}{\partial \epsilon_{ij}^2} = C_{44}^0 - \frac{\lambda^2H_k^2}{a}
\end{eqnarray}
In short, a finite field allows a finite octupole moment, and thus a finite shear strain, to reduce the free energy, reducing the energy cost associated with strain via the $C_{44}^0$ term and thus making the lattice more susceptible to said strain, or softer.

As far as conducting the measurement, a [111] oriented field could be used to measure a combination of $A_5$ and $D_3$ via intermixing the 9 terms in the lower-right quadrant. The $A_6$ coefficient would be induced, but is likely small, as it corresponds to the lowest-order interaction of CEF-forbidden octupoles, or a higher-order interaction invoking CEF-forbidden quadrupoles and dipoles. Alternatively, a [001] aligned magnetic field could be used for measuring a specific elastic constant for the orthogonal shear plane. This, however, would break the degeneracy typical of these three coefficients, inducing the $A_5$ term within only one ($C_{xyxy}$ for $H_z$), meaning the measurement may need to distinguish a newly-differentiated $C_{ijij}$ from the still-equal $C_{jkjk}$ and $C_{ikik}$.

Generally, associated changes in sound velocities/resonant frequencies would likely invoke nearly all $C$ and $D$ coefficients from Table \ref{tab:table2}. However, assuming a [111]-oriented magnetic field, all (field-dependent) contributions associated with the allowed $\Gamma_3$ quadrupoles would cancel (the allowed couplings to field would be to $H_x^2-H_y^2$ and $2H_z^2-H_x^2-H_y^2$). Thus, the remaining coefficients would correspond to CEF-forbidden multipoles, and would likely be small. In contrast, a field aligned along a single principle axis would have a symmetry-allowed coupling to an allowed quadrupole, though the coupling of this quadrupole and the field to shear strains specifically would be higher order and not likely to be significant. Field-independent effects from the $\Gamma_3$ quadrupoles would naturally remain, which would manifest via $C_{11} - C_{12}$, or $C_{11}^{(0)} - C_{12}^{(0)}$ using coefficients from Table \ref{tab:table2}.

Lastly, it should be noted that the $B_1$ coefficients are constrained to be time-reversal odd/imaginary, and thus linear contaminants would likely be either absent or out-of-phase (and thus easily filtered).
\subsection{Magnetic Susceptibility Tensor}

Magnetic susceptibility, herein defined (in slight contrast to convention, and for para-/diamagnetic states) via

\begin{eqnarray}
\chi_{ij}\equiv\left.\frac{\partial^2 F}{\partial H_i \partial H_j}\right|_{H \rightarrow 0} \propto -\frac{\partial M_i}{\partial H_j}=-\frac{\partial M_j}{\partial H_i}
\end{eqnarray}
is a frequently measured quantity, characterizing the linear response of induced magnetic moment to external magnetic field. While the octupole would not produce the simple dipole response typically dominant in susceptibility, the dependence of magnetic susceptibility (quadratically) on strain would give an effective $H\epsilon$ conjugate field and recover $\chi_O\equiv\frac{\partial O}{\partial(H\epsilon)}\propto\frac{\partial^2 F}{\partial(H\epsilon)^2}$, similarly to the aforementioned tensors.

For $O_h$ symmetry, there is a single independent (non-zero) term in the susceptibility tensor in the absence of strain, $\chi_{ii}^0$. Externally induced strains introduce 12 additional independent coefficients (to second order in strain). Thus, for $i\neq j$

\begin{eqnarray}
\chi_{ii} = \chi_{ii}^{(0)}+E\epsilon_{ii}+F(\epsilon_{jj}+\epsilon_{kk})+A_1\epsilon_{ii}^2+\nonumber
\\
A_2(\epsilon_{jj}^2+\epsilon_{kk}^2)+A_3\epsilon_{ii}(\epsilon_{jj}+\epsilon_{kk})+\nonumber
\\
A_4\epsilon_{jj}\epsilon_{kk} + A_6(\epsilon_{ij}^2+\epsilon_{ik}^2)+A_5\epsilon_{jk}^2
\\
\chi_{ij} = G\epsilon_{ij} + D_3\epsilon_{ik}\epsilon_{jk}+D_1\epsilon_{ij}\epsilon_{kk}+ \nonumber
\\
D_2\epsilon_{ij}(\epsilon_{ii}+\epsilon_{jj})
\end{eqnarray}
where $A_5$ and $D_3$ again represent the desired coefficients proportional to the octupole susceptibility, $\frac{\partial^2 F}{\partial(H_i\epsilon_{jk})^2}$ and $\frac{\partial^2 F}{\partial(H_i\epsilon_{jk})\partial(H_j\epsilon_{ik})}$ respectively. As implied by the labeling, many coefficients here are constrained by the definition of the tensors (as derivatives of free energy) to be identical to counterparts in the elastic stiffness tensor.

\subsubsection{Practical Considerations}

Two experimental configurations are suggested. First, to recover the $A_5$ coefficient, susceptibility could be measured along any principal axis, while a shear strain is applied in a plane perpendicular to said axis. The likely application of a net compressive or tensile strain, as opposed to pure shear strain, would induce several other coefficients. The $E$ and $F$ coefficients, in particular, would correspond to allowed bilinear couplings of the $\Gamma_3$ quadrupoles, but are easily experimentally distinguished by their representing linear strain dependencies (as opposed to quadratic). The rest are unlikely to be large, given they do not represent the lowest-order allowed coupling to either allowed quadrupole.

Alternatively, the $D_3$ coefficient could potentially be measured by applying two simultaneous shear strains, and measuring the transverse susceptibility using the two axes perpendicular to said shear strains. In practice, a simpler method would be to use a [111]-aligned magnetic field and a [111] uniaxial stress, inducing all three shear strains simultaneously to measure a combination of $A_5$ and $D_3$. Unfortunately this would likely induce all the coefficients simultaneously, but, again, they would likely be small compared to $A_5$ and $D_3$ given their connection to no multipoles and/or CEF-forbidden multipoles (excepting potentially $E$ and $F$, which would again distinguish themselves from the terms of interest by their linearity in strain).

Many common measurements for magnetic susceptibility involve centering a sample in a detection solenoid and varying field (AC), or setting a field and moving a sample through a detection solenoid (DC), to measure its moment via the response in said solenoid. In either case, unexpected sample movement relative to the detector would generate a spurious signal. Thus, the use of DC strains is motivated, as effects of AC strains would be very difficult to decouple from the effects of sample movements (relative to a detector) that most strain-applying techniques are likely to produce. Unfortunately, this means the susceptibility would have to be measured as a function of strain, with the zero-strain term presenting itself as a constant background; measuring only the strain-dependent term, rather than its sum with the zero-strain susceptibility, would require AC strains. However, with the $\Gamma_3$ doublet being non-magnetic, the strain-independent term should be both generally small and not strongly enhanced by low temperatures, potentially allowing easily-realized strains to drive the octupolar contribution to dominance over any background. Experimental apparatus capable of measuring magnetic moments while compensating for the effects of sample movement, via careful strain application or a detector with significant positional tolerance (perhaps an optical probe or a detector mounted on the strain cell, for instance), may then further apply AC strain and AC magnetic field; an octupole susceptibility could then be isolated from much of the background by measuring the component of the magnetic moment varying with the sum or difference frequency of the strain and magnetic field frequencies.

Lastly, it should be noted that controlling strain would be a potential difficulty, as a measurable octupole susceptibility would lead to a softening of the shear mode with field. Thus, application of constant stress would lead to increasing strain with increasing field. Careful and direct measurement of strain, or the use of a fairly small AC magnetic field for susceptibility measurements, could help mitigate this softening.

\subsection{Non-Linear Magnetic Susceptibility}

While not the primary focus of this paper, the aforementioned $H\epsilon$ product is not the unique lowest-order object the octupole can couple to within the limits of strain and magnetic field; an object of identical symmetry can be constructed simply with a cubic magnetic field term, $H_xH_yH_z$ \footnote{though microscopics may vary, in pure symmetry terms, $\epsilon_{ii}$ and $\epsilon_{ij}$ transform equivalently to $H_i^2$ and $H_iH_j$ respectively; $H_i$ belongs to a $T_1$ representation, and the product of two different field components $H_iH_j$ creates a $T_2$ object symmetrically analogous to a shear strain}. Thus, higher-order magnetization effects can often capture the same information as strain dependencies. Using the same susceptibility definition (albeit without the $H\rightarrow 0$ limit), but expanding in magnetic field rather than in strain, this introduces 5 new independent terms to 4th order; for $i\neq j$,

\begin{eqnarray}
\chi_{ii}^{(0)} + AH_i^2+B(H_j^2 + H_k^2) + \nonumber \\
CH_i^4 + D(H_j^4 + H_k^4) + \nonumber \\
+ 6DH_i^2(H_j^2+H_k^2) + EH_j^2H_k^2 \\
\chi_{ij}= 2BH_iH_j +2EH_iH_jH_k^2 \nonumber\\
+ 4DH_iH_j(H_i^2+H_j^2)
\end{eqnarray}

where the E coefficient represents the desired $\frac{\partial^2 F}{\partial (H_xH_yH_z)^2}$. None of these coefficients are implied by symmetry to be identical to any from the previous tensors.

\subsubsection{Practical Considerations}

Experimentally, the obvious complication is that the high fields potentially necessary to accurately fit a quartic or higher function could render the higher CEF states relevant to the result. Magnetic energy would become comparable to the gap for fields of $\sim$15T-30T depending on the material (likely $\sim$.42T/K for a given CEF gap, which are in the 40-60K range [\onlinecite{Nakatsuji}]).

Two methods present themselves: a simple magnetization-vs-field measurement for a [111]-aligned field and thus [111]-aligned magnetization, and a simple [100] susceptibility measurement with a secondary transverse field along an [011]-type axis. In the [111] case, magnetization would be expected to be $\propto H^5$, or $\frac{dM}{dH} \propto H^4$. Thus, magnetization would have to be sensitively plotted against a fairly wide field range, with a background from the simple dipolar susceptibility being present (but again, likely small for appropriately low field strengths and temperatures, given the CEF splitting). Alternatively, this method would also potentially lend itself to an AC measurement scheme; an AC magnetic field could be applied and the magnetization measured at the fifth harmonic, potentially providing a dramatic improvement in signal-to-background ratio for the octupolar signal.

The alternative [100] case may represent a simpler measurement with a more complicated apparatus. If a strong field could be applied along the [011] axis, a traditional magnetic susceptibility measurement could then be performed along the [100] axis, with the results plotted against $H_{011}$ and fit to a quartic function. Using an AC technique for the [100] susceptibility measurement would eliminate much of the contamination from field misalignment, though background susceptibility from non-octupolar sources would remain a potential issue; in particular, a quadratic dependence on field could potentially arise from a coupling to the $O_2^2$ quadrupole, forbidden with the previous alignment scheme but potentially induced here.

\section{Resistivity}

Resistivity is not a thermodynamic quantity, but terms in the resistivity tensor can nevertheless contain information about the onset of order parameters. Appropriate derivatives of resisitivity tensor elements can then sometimes capture information similar to that in derivatives of the free energy, i.e. thermodynamic probes  [\onlinecite{Max}]. In particular, perturbations that break symmetries of the crystallographic point group can induce changes in resistivity tensor terms, should the perturbation(s) or some product thereof belong to the same irreducible representation as a given resistivity tensor term. Should the applied perturbation also then match the irreducible representation of the order parameter, a term in the change in resistivity will then be linearly proportional to the order parameter, allowing the change in the resistivity to reflect the associated susceptibility to within some proportionality constant. 

Thus, higher rank tensors describing derivatives of resistivity often contain information regarding susceptibility toward symmetry-breaking instabilities, to within some coupling constant. This constant can potentially depend on temperature, or allow certain order parameters to more strongly influence resistivity than others. These complications are generally not insurmountable in extracting the dependence of the underlying order parameter on strain/field. This, combined with the fact that resistivity is, generally, more easily accessed experimentally than many thermodynamic quantities, particularly when trying to measure in a symmetry-selective way, motivates a thorough evaluation.

In the specific context of the $\tau_{xyz}$ octupole allowed in the $\Gamma_3$ doublet, the cyclic permutations of $H_z\epsilon_{xy}$ can couple bilinearly to the octupole, but both of these objects are of $\Gamma_2^+$ symmetry in $O_h$, a symmetry that cannot be constructed purely via elements of the resistivity tensor ($\Gamma_2^+$ has no quadratic basis functions in $O_h$). However, expanding a $\Gamma_5$-type term in the resistivity tensor $\rho_{xy}$, one can note that two objects already present, $H_z$ and $\tau_{xyz}$, together form an object of of appropriate $\Gamma_5$ symmetry. The symmetry-allowed dependency is therefore

\begin{eqnarray}
\Delta\rho_{xy}(H_z,\epsilon_{xy}) \propto H_z\tau_{xyz} \nonumber
\\
\Delta\rho_{xy}(H_z,\epsilon_{xy}) \propto H_z^2\epsilon_{xy}\chi_O
\end{eqnarray}

and thus, the object of relevance is a first derivative with respect to strain and second derivative w.r.t. magnetic field of a resistivity tensor term, i.e. a term in a 6th rank tensor. This object is most easily approached by considering either the second field derivative of the 4th rank elastoresistivity tensor, or the first strain derivative of the 4th rank second-order magnetoresistance tensor. We focus here primarily on the former.

\subsection{Elastoresistivity Tensor}

Elastoresistivity is defined via [\onlinecite{Max}]

\begin{eqnarray}
m_{ij,kl}\equiv\frac{\partial(\frac{\Delta\rho}{\rho})_{ij}}{\partial\epsilon_{kl}}
\end{eqnarray}

Herein the normalized resistivity tensor is defined in 
a manifestly symmetric manner for convenience, $(\frac{\Delta\rho}{\rho}) = \rho^{-1/2}(\Delta\rho)\rho^{-1/2}$[\onlinecite{Max}], enabling the use of the symmetry $(\frac{\Delta\rho}{\rho})_{ij}(H) = (\frac{\Delta\rho}{\rho})_{ji}(-H)$. Thus, the overall tensor is similar, but not identical to, the elastic stiffness tensor; for example, it is not symmetric under exchange of $ij$ and $kl$, and purely dynamic contaminants such as the simple Hall Effect appear in several terms. The full tensor is shown in Table \ref{tab:table3}, to second order in magnetic field; there are only 3 allowed unique field-independent terms, with an additional 15 being induced by applied field. The use of compactified Voigt notation is motivated by this high level of symmetry; excluded terms have identical coefficients to those included on the table, but may have some sign differences, which can be calculated trivially via the symmetries of $\rho_{ij}$ (switching coefficients adds a sign change to each $H$ term) and $\epsilon_{ij}$ (switching coefficients changes nothing); e.g. via the symmetry of $\rho_{ij}$, $m_{zyyy}$ would be $-B_2H_x+D_2H_yH_z$, in slight contrast to $m_{yzyy} = + B_2H_x + D_2H_yH_z$. The $A_6$ (yellow boxes) and $D_5$ (blue boxes) coefficients then represent the desired susceptibility:

\begin{eqnarray}
A_6 = \frac{\partial^2m_{ij,ij}}{\partial H_k^2} \propto \chi_O
\\
D_5 = \frac{\partial^2m_{ij,jk}}{\partial H_k \partial H_i} \propto \chi_O
\end{eqnarray}

It should be further noted that similar notation to previous tensors was chosen for convenience, but that none of these coefficients are constrained by symmetry to have any relationship with those in any of the thermodynamic tensors.

\begin{table*}
\caption{\label{tab:table3}The full elastoresistivity tensor in $O_h$ in compactified Voigt notation, color-coded to indicate which terms have identical or differing coefficients}
\begin{tabular}{ |c|c|c|c|c|c|c| }
\hline
   & $xx$ & $yy$ &$zz$&$yz$&$zx$&$xy$\\
\hline
$xx$ & \cellcolor{red}$m_{11}^{(0)} + A_1H_x^2+$ & \cellcolor{green}$m_{12}^{(0)} + A_3H_z^2+$ & \cellcolor{green}$m_{12}^{(0)} + A_3H_y^2+$&\cellcolor{fuchsia}$D_1H_yH_z$&\cellcolor{tangerine}$D_2H_zH_x$&\cellcolor{tangerine}$D_2H_xH_y$ \\ 
& \cellcolor{red}$A_2(H_y^2+H_z^2)$ & \cellcolor{green}$A_4H_x^2+A_5H_y^2$ & \cellcolor{green}$ A_4H_x^2+A_5H_z^2$&\cellcolor{fuchsia}&\cellcolor{tangerine}& \cellcolor{tangerine}\\ 
\hline
$yy$ &\cellcolor{green}$m_{12}^{(0)}  + A_3H_z^2+$ &\cellcolor{red} $m_{11}^{(0)} + A_1H_y^2+$& \cellcolor{green}$m_{12}^{(0)} + A_3H_x^2+$&\cellcolor{tangerine}$D_2H_yH_z$&\cellcolor{fuchsia}$D_1H_zH_x$&\cellcolor{tangerine}$D_2H_xH_y$ \\
&\cellcolor{green}$ A_4H_y^2+A_5H_x^2$ & \cellcolor{red}$ A_2(H_z^2+H_x^2)$& \cellcolor{green}$A_4H_y^2+A_5H_z^2$&\cellcolor{tangerine}&\cellcolor{fuchsia}&\cellcolor{tangerine} \\
\hline
$zz$ &\cellcolor{green} $m_{12}^{(0)}  + A_3H_y^2+$ & \cellcolor{green}$m_{12}^{(0)} + A_3H_x^2+$&\cellcolor{red}$m_{11}^{(0)} + A_1H_z^2+$&\cellcolor{tangerine}$D_2H_yH_z$&\cellcolor{tangerine}$D_2H_zH_x$&\cellcolor{fuchsia}$D_1H_xH_y$ \\
& \cellcolor{green}$A_4H_z^2+A_5H_x^2$ &\cellcolor{green} $A_4H_z^2+A_5H_y^2$&\cellcolor{red}$A_2(H_x^2+H_y^2)$&\cellcolor{tangerine}&\cellcolor{tangerine}&\cellcolor{fuchsia} \\
\hline
$yz$ & \cellcolor{amethyst}$B_1H_x$&\cellcolor{orange}$B_2H_x$&\cellcolor{orange}$B_2H_x$ & \cellcolor{yellow}$m_{44}^{(0)}+A_6H_x^2+$ & \cellcolor{blue}$B_3H_z$ & \cellcolor{blue}$B_3H_y$ \\
& \cellcolor{amethyst}$+D_3H_yH_z$&\cellcolor{orange}$+D_4H_yH_z$&\cellcolor{orange}$+D_4H_yH_z$ & \cellcolor{yellow}$A_7(H_y^2+H_z^2)$ &\cellcolor{blue} $+D_5H_xH_y$ & \cellcolor{blue}$+D_5H_zH_x$\\
\hline
$zx$ & \cellcolor{orange}$B_2H_y$&\cellcolor{amethyst}$B_1H_y$&\cellcolor{orange}$B_2H_y$ & \cellcolor{blue}$B_3H_z$ & \cellcolor{yellow}$m_{44}^{(0)}+A_6H_y^2+$& \cellcolor{blue}$B_3H_x$ \\ 
& \cellcolor{orange}$+D_4H_zH_x$&\cellcolor{amethyst}$+D_3H_zH_x$&\cellcolor{orange}$+D_4H_zH_x$ & \cellcolor{blue}$+D_5H_xH_y$ &\cellcolor{yellow} $A_7(H_z^2+H_x^2)$ &\cellcolor{blue} $+D_5H_yH_z$\\
\hline
$xy$ & \cellcolor{orange}$B_2H_z$&\cellcolor{orange}$B_2H_z$&\cellcolor{amethyst}$B_1H_z$  & \cellcolor{blue}$B_3H_y$ & \cellcolor{blue}$B_3H_x$ & \cellcolor{yellow}$m_{44}^{(0)}+A_6H_z^2+$  \\ 
& \cellcolor{orange}$+D_4H_xH_y$&\cellcolor{orange}$+D_4H_xH_y$&\cellcolor{amethyst}$+D_3H_xH_y$ & \cellcolor{blue}$+D_5H_zH_x$  &\cellcolor{blue} $+D_5H_yH_z$ & \cellcolor{yellow}$A_7(H_x^2+H_y^2)$\\
\hline
\end{tabular}
\end{table*}

\subsubsection{Practical Considerations}

The tensor presents several obvious experimental opportunities and challenges. First, inspection of the yellow boxes in Table \ref{tab:table3} makes clear that the $m_{xyxy}$ elastoresistivity coefficient is even in $H_z$, and hence that measurement of the $A_6$ coefficient is possible without a linear-in-field contaminant, meaning that it could potentially be extracted as the sole fit parameter of elastoresistivity vs field data. This, in turn, would mean that the coefficient could potentially be extracted with a fairly limited field range, limiting issues arising from high fields (i.e. non-negligible mixing of CEF states). 

Most experimental methods of probing elastoresistivity, however, do not apply pure shear strains, but also induce normal strains $\epsilon_{xx}$, $\epsilon_{yy}$, $\epsilon_{zz}$. The associated symmetry-preserving strain component couples directly to a simple Hall Effect via changing the carrier density; with small strains, charge carrier count would remain constant against an increasing/decreasing volume. Thus, even without a linear-in-field term in the desired $m_{ijij}$ elastoresistivity term, a successful measurement would likely still show a strain-dependent Hall Effect that would need to be accounted for via the traditional methods (this would correspond to an admixture of the $B_1$ and $B_2$ coefficients in the table). For fields aligned precisely along one of the crystal axes $k$, measurement of $\rho_{ij}$ in positive and negative fields would, in principle, allow cancellation of this linear contaminant. Contact misalignment, which can result in admixture of $\rho_{ii}$ in an attempt to measure $\rho_{ij}$, can be subtracted using ideas developed earlier in Ref. ~\onlinecite{Max2}.

Perhaps more importantly, elastoresistivity requires controlling/measuring the strain experienced by a crystal. If an experiment failed to hold strain constant as a function of field, the octupole susceptibility would not be faithfully measured. An example would be the case where stress is held fixed, i.e. a piezoresistance measurement. Given the octupole susceptibility can manifest in the elastic stiffness (see section III-A), the very application of field would change the stiffness independently of temperature, thus changing the strain under conditions of constant stress. Such an effect can be minimized via the use of a strain-applying apparatus that is very stiff relative to the sample, or nearly eliminated by directly measuring and controlling for strain. Appropriate experimental apparatus for such a task have been developed [\onlinecite{Hicks}].

It should be further noted that Table \ref{tab:table3} represents a general compilation of terms allowed in an expansion of resistivity in terms of strain and magnetic field (to first order in strain, second order in field); the order of derivatives is not particularly relevant, and thus strain dependencies of the magnetoresistance would draw from the same set of allowed terms, though high fields (or high strains) would potentially render relevant higher terms than those contemplated here.

\section{Conclusion}

The $\Gamma_3$ doublet ground state for local 4f orbitals in a cubic point symmetry was motivated as an ideal system to study octupole order parameters and their associated susceptibility, given the allowed $\tau_{xyz}$ octupole and the energetic disfavoring of magnetic dipoles. Considering the allowed couplings of such an order parameter, several commonly-measured tensor quantities in which it might appear were discussed. These were fully elucidated in the $O_h$ point group, the point group of experimental realizations of an octupolar order parameter [\onlinecite{Nakatsuji},\onlinecite{Crystal}]. Specific terms within external-field-dependent elastic stiffness, elastoresistivity, and magnetic susceptibility tensors which would be linearly proportional to a potential $\tau_{xyz}$ octupole susceptibility were identified. Potential measurements, and complications arising from contaminant terms, were discussed for each individual tensor, with several octupole-isolating experiments ultimately proposed.

More broadly, similar ideas could be used to isolate contributions of a variety of higher-order local multipoles and in any number of material systems. The chosen system was convenient for both being relatively simple (a doublet ground state) and having no overlap in conjugate fields (the strain component of the octupole conjugate field coupled to no other order parameters allowed by the CEF ground state). Nonetheless, the core idea of isolating specific multipolar contributions to potentially rich phase diagrams via higher-rank tensor properties is broadly applicable to a variety of localized 4f systems.

\subsection{Acknowledgements}

The authors would like to thank M.C. Shapiro, E.W. Rosenberg, R.M. Fernandes, and B.J. Ramshaw for helpful conversations. We particularly thank R.M. Fernandes for pointing out the possibility to measure the octupole susceptibility via the 5th harmonic of the AC magnetic susceptibility. This work was supported by the Gordon and Betty Moore Foundation EPiQS Initiative, through Grant GBMF9068. M.E.S. was supported by a NSF Graduate Research Fellowship (Grant No. DGE-114747).

\appendix
\section{Character Table}

See Table \ref{tab:Oh}. For convenience, a series of symmetrized cubic rotation products have been added. These have the same spatial symmetries as the magnetic octupole, and thus indicate the irreducible representations of the various possible magnetic octupole moments.

\begin{table*}
    \centering
    \begin{tabular}{c|c|c|c|c|c|c|c|c|c|c|c|c|c|c}
    $O_h$& & $E$ & $8C_3$ & $6C_2$ & $6C_4$ & $3(C_4)^2$ & $i$ & $6S_4$ & $8S_6$ & $3\sigma_h$ & $6\sigma_d$ & linear functions & quadratic functions & cubic functions and\\ 
    &&&&&&&&&&&&and rotations&&cubic rotation products\\ \hline
    $\Gamma_1^+$ & $A_{1g}$ & +1 & +1 & +1 & +1 & +1 & +1 & +1 & +1 & +1 & +1 & - & $x^2+y^2+z^2$ & - \\ \hline
    $\Gamma_2^+$ & $A_{2g}$ & +1 & +1 & -1 & -1 & +1 & +1 & -1 & +1 & +1 & -1 & - & - & $\overline{R_xR_yR_z}$\\ \hline
    $\Gamma_3^+$ & $E_{g}$ & +2 & -1 & 0 & 0 & +2 & +2 & 0 & -1 & +2 & 0 & - & $(2z^2-x^2-y^2,$ & - \\
    &&&&&&&&&&&&&$x^2-y^2)$&\\ \hline
    $\Gamma_4^+$ & $T_{1g}$ & +3 & 0 & -1 & +1 & -1 & +3 & -1 & 0 & -1 & -1 & $(R_x,R_y,R_z)$ & - & $(R_x^3,R_y^3,R_z^3)$ \\ 
    & & & & & & & & & & & & & & $(\overline{R_xR_z^2}+\overline{R_xR_y^2},\overline{R_yR_x^2}+\overline{R_yR_z^2},$ \\ 
    & & & & & & & & & & & & & & $\overline{R_zR_y^2}+\overline{R_zR_x^2})$ \\ \hline
    $\Gamma_5^+$ & $T_{2g}$ & +3 & 0 & +1 & -1 & -1 & +3 & -1 & 0 & -1 & +1 & - & $(yz,zx,xy)$ & $(\overline{R_xR_z^2}-\overline{R_xR_y^2},\overline{R_yR_x^2}-\overline{R_yR_z^2},$ \\ 
    & & & & & & & & & & & & & & $\overline{R_zR_y^2}-\overline{R_zR_x^2})$ \\ \hline
    $\Gamma_1^-$ & $A_{1u}$ & +1 & +1 & +1 & +1 & +1 & -1 & -1 & -1 & -1 & -1 & - & - & - \\ \hline
    $\Gamma_2^-$ & $A_{2u}$ & +1 & +1 & -1 & -1 & +1 & -1 & +1 & -1 & -1 & +1 & - & - & xyz \\ \hline
    $\Gamma_3^-$ & $E_{u}$ & +2 & -1 & 0 & 0 & +2 & -2 & 0 & +1 & -2 & 0 & - & - & - \\ \hline
    $\Gamma_4^-$ & $T_{1u}$ & +3 & 0 & -1 & +1 & -1 & -3 & -1 & 0 & +1 & +1 & $(x,y,z)$ & - & $(x^3,y^3,z^3)$ \\
    &&&&&&&&&&&&&&$(xz^2+xy^2,yx^2+yz^2,zy^2+zx^2)$\\ \hline
    $\Gamma_5^-$ & $T_{2u}$ & +3 & 0 & +1 & -1 & -1 & -3 & +1 & 0 & +1 & -1 & - & - & $(xz^2-xy^2,yx^2-yz^2,zy^2-zx^2)$ \\ \hline
    \end{tabular}
    \caption{$O_h$ Character Table}
    \label{tab:Oh}
\end{table*}

\section{Table Symmetries}

Herein, terms are defined by "types," where a given type is defined by having a unique index composition (i.e. $ii,ii$ vs $ii,jj$), and $i\neq j\neq k$ holds for all types.
A type then constitutes a term and all terms that can be generated from that (arbitrary) original term by various symmetries, which can be simplified to include only the symmetries of a given tensor, the 3-fold rotational symmetry, and the various 4-fold rotations. For example, Type II for the elastic stiffness tensor, $C_{ii,jj}$, includes $C_{xx,yy}$, $C_{yy,xx}$ (owing to the symmetry of the tensor; see relevant section below), $C_{yy,zz}$, etc. The wording "sign change" is used to indicate the operation $(x)\rightarrow(-x)$ for a given variable a tensor depends on, such as magnetic field.

\subsection{Elastic Stiffness}

Here the $C$ symmetry is defined as that which exchanges the two subsets of indices (i.e. $C_{ab,cd}\rightarrow C_{cd,ab}$), while the $\epsilon$ symmetry is defined as that which switches indices within a subset ($C_{ab,cd}\rightarrow C_{ba,cd}$).

\textbf{Type I}: $C_{ii,ii}$ (Red Boxes in Table \ref{tab:table2})

1. Invariant under simultaneous sign change of any two field components ($\sigma_i$/$\sigma_j$/$\sigma_k$)

2. Invariant under simultaneous exchange of $H_j$ and $H_k$ and sign change of $H_i$ ($\sigma_{j=k}$)

Final Form: $C_{11}^0+A_1H_i^2+A_2(H_j^2+H_k^2)$

\textbf{Type II}: $C_{ii,jj}$ (Green Boxes in Table \ref{tab:table2})

1. Invariant under simultaneous sign change of any two field components ($\sigma_i$/$\sigma_j$/$\sigma_k$)

2. Invariant under simultaneous exchange of $H_i$ and $H_j$ and sign change of $H_k$ ($\sigma_{i=j}$,$C$)

Final Form: $C_{12}^0 + A_3(H_i^2+H_j^2)+A_4(H_k^2)$

\textbf{Type III}: $C_{ij,ij}$ (Yellow Boxes in Table \ref{tab:table2})

1. Invariant under simultaneous sign change of any two field components ($\sigma_i$/$\sigma_j$/$\sigma_k$)

2. Invariant under simultaneous exchange of $H_i$ and $H_j$ and sign change of $H_k$ ($\sigma_{i=j}$,$\epsilon$)

Final Form: $C_{44}^0 + A_6(H_i^2+H_j^2)+A_5(H_k^2)$

\textbf{Type IV}: $C_{ii,ij}$ (Orange Boxes in Table \ref{tab:table2})

1. Zero in the absence of symmetry-breaking field, magnetic or otherwise ($\sigma_i$ or $\sigma_j$)

2. Antisymmetric under simultaneous sign change of $H_k$ and $H_j$/$H_i$ ($\sigma_i$/$\sigma_j$)

Final Form: $B_1H_k+D_2H_iH_j$ 

\textbf{Type V}: $C_{ij,kk}$ (Purple Boxes in Table \ref{tab:table2})

1. Zero in the absence of symmetry-breaking field, magnetic or otherwise ($\sigma_i$ or $\sigma_j$)

2. Invariant under simultaneous exchange of $H_i$ and $H_j$ and sign change of $H_k$ ($\sigma_{j=k}$,$\epsilon$)

3. Antisymmetric under exchange of $H_i$ and $H_j$ followed by sign change of $H_i$ ($C_{4k}$,$\epsilon$)

Final Form: $D_1H_iH_k$

\textbf{Type VI}: $C_{ij,jk}$ (Blue Boxes in Table \ref{tab:table2})

1. Zero in the absence of symmetry-breaking field, magnetic or otherwise ($\sigma_i$ or $\sigma_k$)

2. Invariant under simultaneous exchange of $H_i$ and $H_k$ and sign change of $H_j$ ($\sigma_{i=k}$,$C$,$\epsilon$)

3. Antisymmetric under exchange of $H_i$ and $H_k$ followed by sign change of $H_i$ ($C_{4j}$, $C$)

Final Form: $D_3H_iH_k$

\subsection{Strain-dependent Magnetic Susceptibility}

The magnetic susceptibility tensor, again defined by 

\begin{eqnarray}
\chi_{ij}\equiv\left.\frac{\partial^2 F}{\partial H_i \partial H_j}\right|_{H \rightarrow 0} \propto -\frac{\partial M_i}{\partial H_j}=-\frac{\partial M_j}{\partial H_i}
\end{eqnarray}

has one obvious symmetry. This symmetry, herein defined as "$\chi$" symmetry, implies invariance under simple exchange of indices, i.e. $\chi_{ij}\rightarrow \chi_{ji}$

\textbf{Type I}: $\chi_{ii}$

1. Invariant under sign-change of $i$/$j$/$k$ indices ($\sigma_i$/$\sigma_j$/$\sigma_k$)

2. Symmetric under exchange of $j$ and $k$ indices ($\sigma_{j=-k}$)

Final Form: $\chi_{ii} = \chi_{ii}^{(0)}+A\epsilon_{ii}+B(\epsilon_{jj}+\epsilon_{kk})+C\epsilon_{ii}^2+\nonumber
\\
D(\epsilon_{jj}^2+\epsilon_{kk}^2)+E\epsilon_{ii}(\epsilon_{jj}+\epsilon_{kk})+\nonumber
\\
F\epsilon_{jj}\epsilon_{kk} + G(\epsilon_{ij}^2+\epsilon_{ik}^2)+L\epsilon_{jk}^2$

\textbf{Type II}: $\chi_{ij}$

1. Zero in the absence of symmetry-breaking field, strain or otherwise ($\sigma_i$ or $\sigma_j$)

2. Antisymmetric under sign change of $i$/$j$ ($\sigma_i$/$\sigma_j$)

3. Invariant under exchange of $i$ and $j$ coefficients ($\sigma_{i=-j}$. $\chi$)

Final Form: $M\epsilon_{ij} + N\epsilon_{ik}\epsilon_{jk}+O\epsilon_{ij}\epsilon_{kk}+P\epsilon_{ij}(\epsilon_{ii}+\epsilon_{jj})$

It can then be noted that, given the definition of $C$ and the definition of $\chi$, each of these terms corresponds to some allowed term in the free energy, and the terms which give rise to many of the $C$ tensor terms are identical to many that give rise to the strain-dependent $\chi$ tensor terms. Thus, the terms can be rewritten as

Final Form: $\chi_{ii} = \chi_{ii}^{(0)}+E\epsilon_{ii}+F(\epsilon_{jj}+\epsilon_{kk})+A_1\epsilon_{ii}^2+\nonumber
\\
A_2(\epsilon_{jj}^2+\epsilon_{kk}^2)+A_3\epsilon_{ii}(\epsilon_{jj}+\epsilon_{kk})+\nonumber
\\
A_4\epsilon_{jj}\epsilon_{kk} + A_6(\epsilon_{ij}^2+\epsilon_{ik}^2)+A_5\epsilon_{jk}^2$

Final Form: $G\epsilon_{ij} + D_3\epsilon_{ik}\epsilon_{jk}+D_1\epsilon_{ij}\epsilon_{kk}+D_2\epsilon_{ij}(\epsilon_{ii}+\epsilon_{jj})$

\subsection{Non-linear Magnetic Susceptibility}

The inherent symmetry of the tensor here remains $\chi_{ij} \rightarrow \chi_{ji}$, as in the previous case.

\textbf{Type I}:$\chi_{ii}$

1. Invariant under simultaneous sign change of any two field components ($\sigma_i$/$\sigma_j$/$\sigma_k$)

2. Invariant under simultaneous exchange of $H_j$,$H_k$ and sign change of $H_i$ ($\sigma_{j=-k}$)

Final Form: $\chi_{ii}^{(0)} + AH_i^2+B(H_j^2 + H_k^2) + \nonumber \\
CH_i^4 + D(H_j^4 + H_k^4) + \nonumber \\
+ EH_i^2(H_j^2+H_k^2) + FH_j^2H_k^2$

\textbf{Type II}:$\chi_{ij}$

1. Zero in the absence of symmetry-breaking field, magnetic or otherwise ($\sigma_i$ or $\sigma_j$)

2. Invariant under simultaneous sign change of $H_i$,$H_j$ ($\sigma_k$)

2. Antisymmetric under simultaneous sign change of $H_j$,$H_k$/$H_k$,$H_i$ ($\sigma_{i}$/$\sigma_{j}$)

3. Invariant under simultaneous exchange of $H_i$,$H_j$ and sign change of $H_k$ ($\sigma_{i=-j}$,$\chi$)

Final Form: $\chi_{ij}= GH_iH_j +LH_iH_jH_k^2
\\
+ NH_iH_j(H_i^2+H_j^2) + OH_k(H_i^2-H_j^2)$

Furthermore, moving beyond Neumann's Principle, it can be noted that the aforementioned definition of magnetic susceptibility implies each term derives from a corresponding term in the free energy. Some of these $\chi$ tensor terms are then implied to derive from the same allowed term within the free energy, and are thus constrained to be equal, to within a numerical factor (given different derivative orders). Additionally, one term allowed by Neumann's principle in $\chi_{ij}$, $OH_k(H_i^2-H_j^2)$, implies a term in $\chi_{ii}$, $H_iH_jH_k$, that is forbidden, and is thus not allowed (alternatively, the free-energy term implied by $OH_k(H_i^2-H_j^2)$ is found to cancel if the equivalent free-energy terms from $\chi_{ij}$,$\chi_{jk}$, and $\chi_{ki}$ are added together, yielding $H_iH_jH_k(H_i^2-H_j^2+H_j^2-H_k^2+H_k^2-H_i^2$).

Thus, the allowed terms can be further simplified to:

$\chi_{ii}^{(0)} + AH_i^2+B(H_j^2 + H_k^2) + \nonumber \\
CH_i^4 + D(H_j^4 + H_k^4) + \nonumber \\
+ 6DH_i^2(H_j^2+H_k^2) + EH_j^2H_k^2$

$\chi_{ij}= 2BH_iH_j +2EH_iH_jH_k^2
\\
+ 4DH_iH_j(H_i^2+H_j^2)$

\subsection{Elastoresistivity}

Elastoresistivity, defined again by
\begin{eqnarray}
m_{ij,kl}\equiv\frac{\partial(\frac{\Delta\rho}{\rho})_{ij}}{\partial\epsilon_{kl}} \nonumber
\end{eqnarray}
does not admit the exchange of the index pairs, i.e. $m_{ij,kl} \rightarrow m_{kl,ij}$. Thus, the symmetries of the constituent components are the only major symmetries of the tensor itself. First, the inherent "$\epsilon$" symmetry implies invariance under $m_{ij,kl}\rightarrow m_{ij,lk}$. Next, the symmetry of the normalized resistivity tensor, defined here (for the purposes of symmetry[\onlinecite{Max}]) via 
\begin{eqnarray}
(\frac{\Delta\rho}{\rho}) = \rho^{-1/2}(\Delta\rho)\rho^{-1/2}
\end{eqnarray}
implies invariance under the "$\rho$" symmetry operation, $m_{ij,kl} \rightarrow -m_{ji,kl}$, as noted in the relevant section above.

\textbf{Type I}: $m_{ii,ii}$ (Red Boxes in Table \ref{tab:table3})

1. Even in $H_i$/$H_j$/$H_k$ ($\sigma_i$/$\sigma_j$/$\sigma_k$,$\rho$)

2. Invariant under exchange of $H_j$,$H_k$ ($\sigma_{j=k}$,$\rho$)

Final Form: $m_{11}^0 + A_1H_i^2+A_2(H_j^2+H_k^2)$

\textbf{Type II}: $m_{ii,jj}$ (Green Boxes in Table \ref{tab:table3})

1. Even in $H_i$/$H_j$/$H_k$ ($\sigma_i$/$\sigma_j$/$\sigma_k$,$\rho$)

Final Form: $m_{12}^0 + A_3H_k^2+A_4H_i^2+A_5H_j^2$

\textbf{Type III}: $m_{ij,ij}$ (Yellow Boxes in Table \ref{tab:table3})

1. Invariant under simultaneous sign change of any two field components ($\sigma_i$/$\sigma_j$/$\sigma_k$)

2. Invariant under exchange of $H_i$,$H_j$ ($\sigma_{i=-j}$,$\rho$,$\epsilon$)

Final Form: $m_{44}^0 + A_6H_k^2+A_7(H_i^2+H_j^2)$

\textbf{Type IV}: $m_{ii,ij}$ (Peach Boxes in Table \ref{tab:table3})

1. Zero in the absence of symmetry-breaking field, magnetic or otherwise ($\sigma_i$ or $\sigma_j$)

2. Antisymmetric under simultaneous sign change of $H_j$,$H_k$/$H_k$,$H_i$ ($\sigma_{i}$/$\sigma_{j}$)

3. Invariant under simultaneous sign change of $H_i$,$H_j$ ($\sigma_k$)

4. Invariant under simultaneous sign change of $H_i$,$H_j$,$H_k$ ($\rho$)

Final Form: $D_2H_iH_j$

\textbf{Type V}: $m_{ij,jj}$ (Orange Boxes in Table \ref{tab:table3})

1. Zero in the absence of symmetry-breaking field, magnetic or otherwise ($\sigma_i$ or $\sigma_j$)

2. Antisymmetric under simultaneous sign change of $H_j$,$H_k$/$H_k$,$H_i$ ($\sigma_{i}$/$\sigma_{j}$)

3. Invariant under simultaneous sign change of $H_i$,$H_j$ ($\sigma_k$)

Final Form: $B_2H_k+D_4H_iH_j$

\textbf{Type VI}:$m_{ij,kk}$ (Purple Boxes in Table \ref{tab:table3})

1. Zero in the absence of symmetry-breaking field, magnetic or otherwise ($\sigma_i$ or $\sigma_j$)

2. Antisymmetric under simultaneous sign change of $H_j$,$H_k$/$H_k$,$H_i$ ($\sigma_{i}$/$\sigma_{j}$)

3. Invariant under simultaneous sign change of $H_i$,$H_j$ ($\sigma_k$)

4. Invariant under exchange of $H_i$,$H_j$ ($\sigma_{x=-y}$,$\rho$)

Final Form: $B_1H_k + D_3H_iH_j$

\textbf{Type VII}: $m_{ii,jk}$ (Violet Boxes in Table \ref{tab:table3})

1. Zero in the absence of symmetry-breaking field, magnetic or otherwise ($\sigma_j$ or $\sigma_k$)

2. Odd in $H_j$/$H_k$ ($\sigma_j$/$\sigma_k$,$\rho$)

3. Even in $H_i$ ($\sigma_i$,$\rho$)

Final Form:$D_1H_jH_k$

\textbf{Class VIII}: $m_{ij,jk}$ (Blue Boxes in Table \ref{tab:table3})

1. Zero in the absence of symmetry-breaking field, magnetic or otherwise ($\sigma_i$ or $\sigma_k$)

2. Antisymmetric under simultaneous sign change of $H_j$,$H_k$/$H_i$,$H_j$ ($\sigma_{i}$/$\sigma_{k}$)

3. Invariant under simultaneous sign change of $H_i$,$H_k$ ($\sigma_j$)

Final Form: $B_3H_j + D_5H_kH_i$

\bibliography{Gamma3}

\end{document}